\documentclass{article}
\usepackage{graphicx} 
\usepackage[letterpaper, margin=1in]{geometry}
\usepackage[numbers, sort, compress]{natbib}
\usepackage{amsmath}
\usepackage{caption}
\title{Low Speed Oblique Impact Behavior On Granular Media Across Gravitational Conditions; The role of cohesion}

\author{
  Seungju Yeo$^1$, Rachel Glade$^{2,1}$, Alice Quillen$^3$, Hesam Askari$^1$
  \vspace{1em}
  \\
  1. Department of Mechanical Engineering,
  University of Rochester,
  Rochester NY, 14627, USA\\
  2. Department of Earth and Environmental Sciences,
  University of Rochester,
  Rochester NY, 14627, USA\\
  3. Department of Physics and Astronomy,
  University of Rochester,
  Rochester NY, 14627, USA\\
  }

\begin{document}

\begin{center}
    \Large Low Speed Oblique Impact Behavior On Granular Media Across Gravitational Conditions; The role of cohesion\\
\vspace{1em}
    \large
    Seungju Yeo$^1$, Rachel Glade$^{2,1}$, Alice Quillen$^3$, Hesam Askari$^1$
  \vspace{1em}
  \\
  1. Department of Mechanical Engineering,
  University of Rochester,
  Rochester NY, 14627, USA\\
  2. Department of Earth and Environmental Sciences,
  University of Rochester,
  Rochester NY, 14627, USA\\
  3. Department of Physics and Astronomy,
  University of Rochester,
  Rochester NY, 14627, USA\\
  
\end{center}

\section*{Abstract}

Analyses of impact provide rich insights from the evolution of granular bodies to their structural properties of the surface and subsurface layers of celestial bodies.
Although chemical cohesive bonding has been observed in asteroid samples, and low-speed impact has been a subject of many studies, our understanding of the role of cohesion in these dynamics is limited, especially at small gravities such as those observed on asteroid surfaces.
In this work, we use numerical discrete element method (DEM) and analytical dynamic resistive force theory (DRFT) modeling to examine the effect of cohesion on the outcome of the impact into loose granular media and explore scaling laws that predict impact behavior in the presence of cohesion under various gravitational conditions and cohesive strengths.
We find that the effect of cohesion on the impact behavior becomes more significant in smaller gravitational acceleration, raising the need to scale the cohesion coefficient with gravity.
We find that due to an insufficient understanding of confounding between cohesion and friction-induced quasi-static and inertial resistance, the outcomes of the DEM simulation models are incongruent with a suggested analytic model using Froude and Bond number scaling based on an additive contribution of frictional and inertial forces. Our study suggests that new dimensionless parameters and scaling are required to accurately capture the role of cohesion, given its ties to frictional behavior between the grain particles at different gravities.

\section{Introduction}
\label{sec:intro}
The impact of foreign objects onto celestial bodies is a common occurrence in our solar system.
Nevertheless, impacts provide rich information about the history, properties, and composition of planetary bodies.
For example, the depth of the subsurface liquid layer below the craters on Ganymede and Callisto was estimated from the impact features of the craters \cite{WHITE2025}. 
In another work, observational data of the craters of the Moon were examined to find pre-mare conditions, before a series of impacts that formed the mare \cite{mare2024}.
It was also shown that the bulk density of porous asteroids increases over time as a result of surface impacts \cite{asteroid_impact2023}, underscoring the role of impacts in the surface evolution of planetary bodies.
These instances show how impact studies provide a deeper understanding of the structure and the evolution history of planetary bodies and natural satellites.
\par
Recent studies have examined specific aspects of impacts such as the geometry of the crater, the number of ejecta, the kinetic energy distribution of the ejected particles or post-impact impactor behavior \cite{esteban2020,Ralaiarisoa2022,Beladjine_2007,rioual2000,valance2009,HOUSENHOLSAPPLE2011,cline2022}.
These quantified measurands of impact are dependent on several factors such as impact speed, projectile, and target properties.
The speed regime of impact heavily influences impact behavior and crater formation; as \cite{highspeed1960} demonstrated, a high-speed impact of a steel ball on blocks of soft lead has a well-defined crater compared to a low-speed impact. 
Along with speed, the properties of the target body affect impact behavior. In particular, the role of cohesion between the grains is a major factor.  This is demonstrated by examination of the impact dynamics and crater formation morphology between cohesive rocks and loose granular material \cite{guldemeister2015scaling}.
In another work, utilizing different levels of cohesion within the concrete binder sample resulted in different impact behavior in the study of the effect of thermal aging \cite{YOURONG2019}.
They showed that cohesive strength increases the elasticity of a surface and thus affects the initial crater morphology upon impact and post-impact stages.
The same applies to granular surfaces composed of discrete grains, regardless of the mechanism of cohesion - capillary, electrostatic, or van der Waals force \cite{Israelachvili2011}.\par
Therefore, one can further infer that impact characteristics are affected by cohesion because cohesion alters the surface elasticity and cratering dynamics.
For instance, impacts on a cohesive granular media eject fewer particles than those on a cohesion-less media \cite{Ralaiarisoa2022}.
Observations show that cohesion counteracts gravity on the size of the crater in high-speed impact experiments under micro-gravity conditions \cite{KIUCHI2023}. The connection between gravity and cohesion is often made by using Bond number as a ration of ${\rho g a^2}/{\gamma_s}$, where $a$ is a characteristic length, $\rho$ is the density of grains, $g$ is the acceleration of gravity, and $\gamma_s$ is the surface energy density driving the cohesive forces.
Froude number and Bond number are used to connect gravitational conditions and cohesive strengths of the impacted granular bodies in recent works \cite{Ralaiarisoa2022,KIUCHI2023}.
However, the combined effect of cohesion and gravity in low-speed impact dynamics under micro-gravity conditions is less studied.\par
For the purpose of establishing a physical relationship between cohesion, gravity, and impact behavior into granular granular surfaces, this study examines the effect of cohesion under various gravitational conditions.
A physical equation is suggested to relate acceleration of gravity and cohesion so that the level of cohesion of the granular surface of planetary bodies can be acquired from the impact trajectory and the crater. 
To achieve the aim, we simulate the impact trajectory of a disc onto a 2D granular bed consisting of polydisperse spherical grains under different cohesion and gravity conditions with varying impact angles and velocities using Large-scale Atomic/Molecular Massively Parallel Simulator (LAMMPS) \cite{LAMMPS}. 
This study assists design of landing missions which include low-speed collision into granular media under micro-gravity conditions, such as asteroids with unconsolidated surface grains of Itokawa, Ryugu, and Bennu \cite{tagsam2022,itokawasampling2006,Philaesampling2020,Itokawa_sample2007,ryugu_sample2023,bennu_sample_return2024}, and many more due to its shock-absorbing nature \cite{rubble_pile_forever2023}.
\par

\section{Methods} 
For the purpose of studying granular material, the Discrete Element Method (DEM) is applied to compute grain kinematics over temporal increments, assuming each grain is a deformable soft-disc with Hertzian contacts.
The repulsive forces, $\hat{\mathbf{F}}$, in normal direction of the contact, $\hat{\mathbf{n}}$, among those soft-discs are proportional to the level of overlap between the grains as shown in the first term of Eq. \ref{eq:hertzcoh}:\par
\begin{align}
\hat{\mathbf{F}} = \left(\frac{4}{3}E_{eff}R^{1/2}_{eff}\delta^{3/2}_{ij}-4\pi\gamma_s R_{eff}-\eta_{n0} m_{eff}v_{ij} \right)\hat{\mathbf{n}}
\label{eq:hertzcoh}
\end{align}
where \(E_{eff}=((1-\nu_i^2)/E_i+(1-\nu_j^2)/{E_j})^{-1}\), \(R_{eff}=R_iR_j/(R_i+R_j)\) and \(\delta_{ij}\) is the overlap for particles $i$ and $j$.
The second term defines the cohesive force characterized by surface energy density $\gamma_s$ in $[J m^{-2}]$, \cite{DERJAGUIN1975}.
The third term represents mass-velocity damping model among the particles where \(\eta_{n0}\) is a damping coefficient for a normal contact, \(m_{eff}=m_im_j/(m_i+m_j)\) and \(v_{ij}\) is the component of relative velocity along $\hat{\mathbf{n}}$.
The value of \(\eta_{n0}\) is determined empirically to match experimental results, as further discussed later in this section.  
\par
\begin{figure}[ht]
\centering
\includegraphics[width=0.4\textwidth]{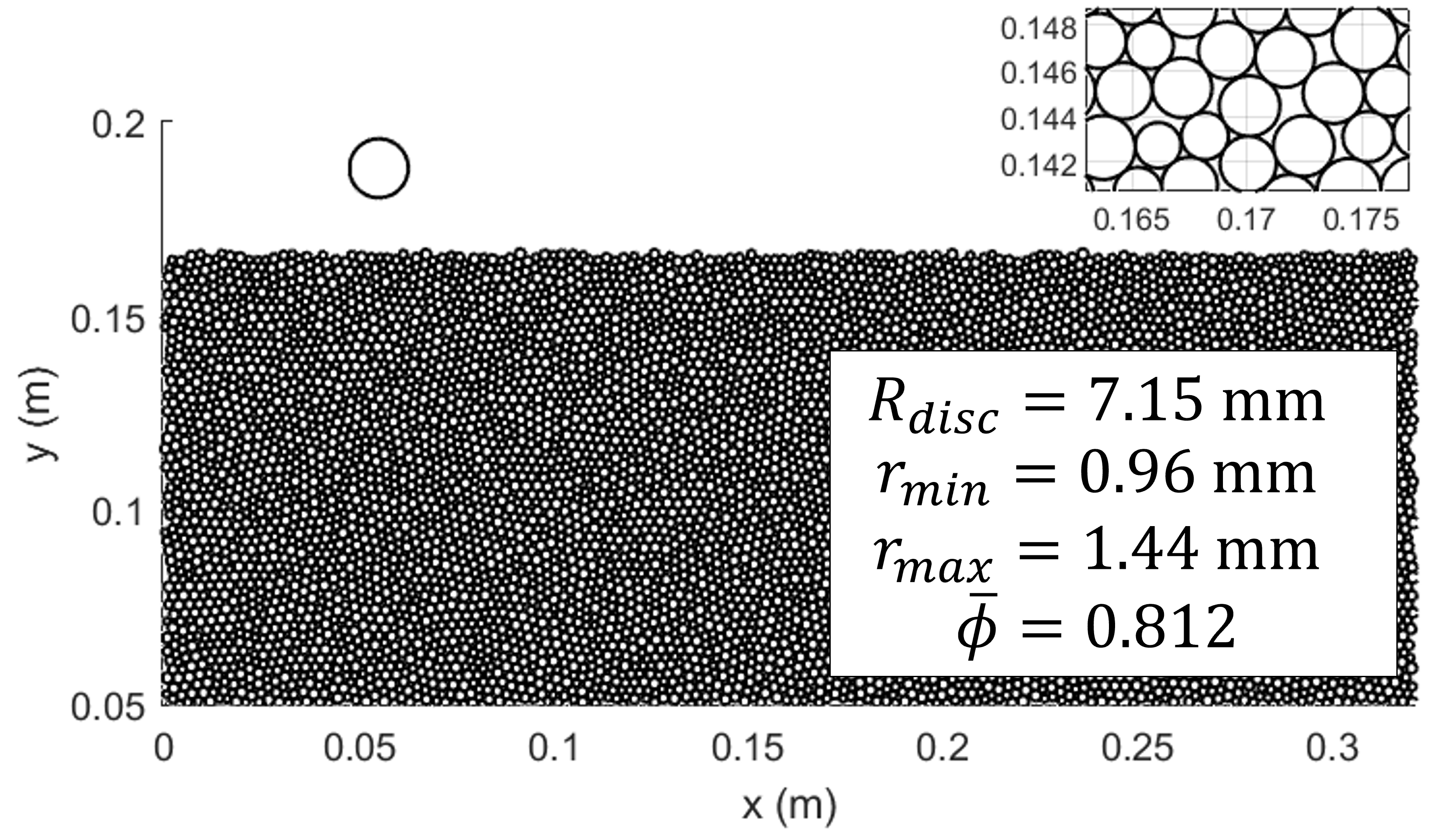}
\caption{Initial configuration of granular bed settled under Earth gravity. The impact disc with radius $a$=7.58 mm is shown above the granular surface at (0.055, 0.188). An enlarged view is shown as an inset on the top right. The minimum grain radius is $r_{min}$=0.96 mm, and the maximum grain radius is $r_{max}$=1.44 mm with a uniform distribution. The average packing fraction $\phi$=0.812.}
\label{fig:bedconfiguration}
\end{figure}
\begin{figure}[ht]
\centering
\includegraphics[width=0.4\textwidth]{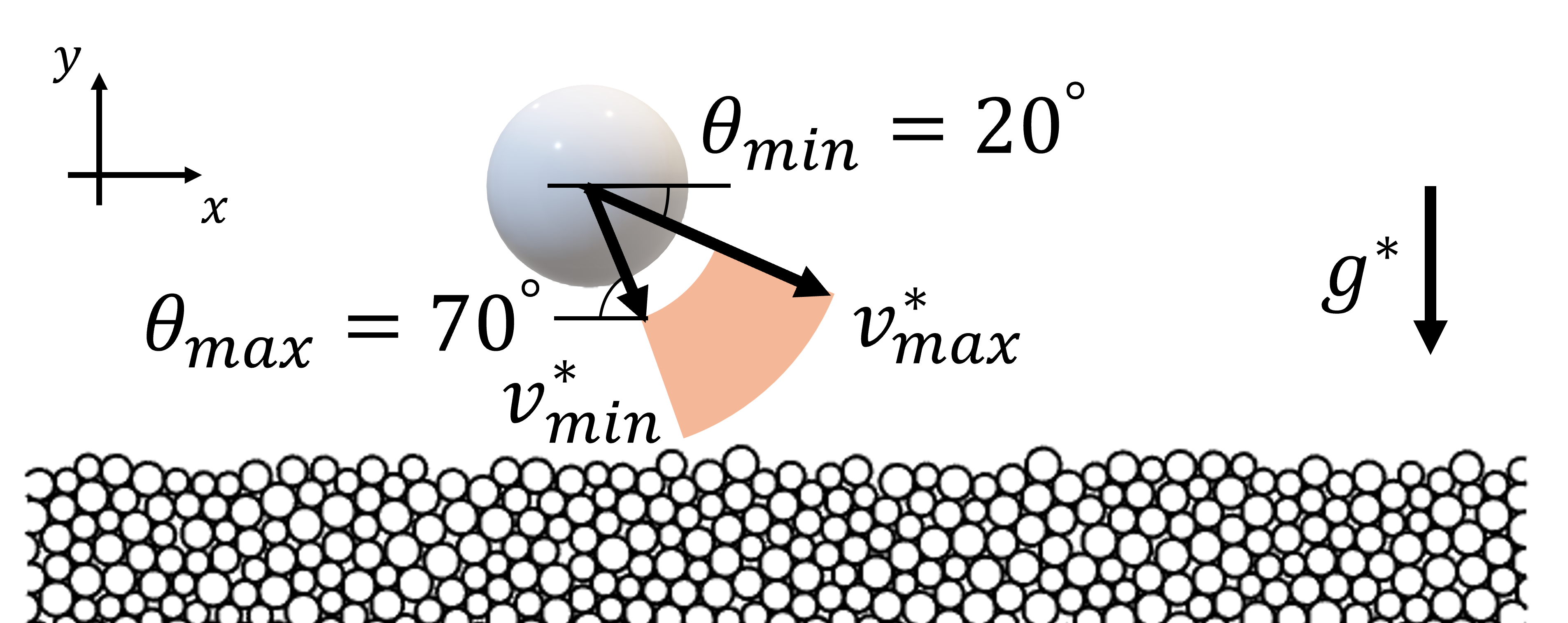}
\caption{Convention of impact condition. The angle of the initial velocity varies from 20$^\circ$ to 70$^\circ$, with respect to the horizon. The speed is scaled with Froude number (i.e. $v^*_{min}=1$m s$^{-1} \times \sqrt{g^*/g_E}$ and $v^*_{max}=7$m s$^{-1} \times \sqrt{g^*/g_E}$ where $g_E$ and $g^*$ are the Earth and the corresponding gravity constant, respectively).}
\label{fig:impact_convention}
\end{figure}

We simulate oblique impacts approximately  6900 polydisperse discs under friction, cohesion, and gravitational forces.
As shown in Figure \ref{fig:bedconfiguration} and \ref{fig:impact_convention}, grain particles with mean radius $\bar{r}$=1.20 $\pm$ 0.24 mm form a bed measuring about 133$\bar{r}\times$70$\bar{r}$ with a rough free surface at the top.
The particles are poured down from a height of about 350$\bar{r}$ above the bottom wall under Earth gravity, and settle down for 6 simulation seconds to dampen grain vibrations.\par
The gravity constant of the simulation is then gradually reduced to Moon level for over 1 simulation second and then to Bennu level for over 10 simulation seconds, given the larger order of magnitude difference between Bennu and Moon conditions.
The purpose of such gradual reduction of gravity is to minimize difference of bed configuration among different gravity levels.\par
After the reduction of gravity, the bed is relaxed for 1 and 70 simulation seconds for Moon and Bennu, respectively, to ensure the residual motion is subsidized.
The bed has a fixed boundary in vertical direction, with few rows of particles below the height of 0.06 m to prevent consolidation upon shear, and a periodic boundary in horizontal direction.\par
The overall packing fraction of the bed configuration at the initial step is found to be 0.83, 0.82, and 0.82 for Earth, Moon, and Bennu gravity conditions, respectively, based on the Delaunay triangulation method.
Even though the distribution of the particles remains the same, the overlap between particles decreases at a weaker gravity field due to the reduction of the weight of grains, which results in a slightly more porous granular bed.\par
The impactor disc of radius $a$= 7.58 mm is placed about 2 cm above the surface, shot into the bed with a velocity ranging from 1.0 m s$^{-1}$ to 7.0 m s$^{-1}$ in 0.5 m s$^{-1}$ increments and at an angle varied from 20$^\circ$ to 70$^\circ$ in 5$^\circ$ increments from the surface horizon.
These range and increments of the velocity and angle yield 143 cases of simulation for each bed with corresponding gravity condition and cohesion.\par
In this study, 12 different beds are explored using three gravity conditions - Earth, Moon, and Bennu - and four cohesion values, thus exploring 1716 simulations in total.
The selected cohesion values are addressed later in the same section.
For Moon and Bennu gravity conditions (Table \ref{tab:gravity}), impact velocity is accordingly scaled with Froude number $Fr=v/\sqrt{ga}$ while range and increment of the impact angle and the initial location of the impact disc are consistent across different gravity conditions \cite{peter2023}.
However, since the pre-impact location of the impactor is fixed, the collision point between the impactor disc and the surface is inconsistent among different angles at the same impact velocity.
Such inconsistency counterbalances the effect of the local packing fraction and the surface irregularity on the impact behavior.
To further neutralize the localized effects due to the random configuration of the grains at the impact site, the impacts are repeated at three different initial impactor locations.
\par


\begin{table}[ht]
    \centering
    \caption{Gravitational acceleration constant for different planetary bodies}
    \label{tab:gravity}
    \begin{tabular}{ll}
    Planetary Body & Gravitational acceleration ($m s^{-2}$) \\
    \hline
    Earth & $9.81$ \\
    Moon & $1.625$\\
    Bennu & $6.27E-5$\\
\hline
    \end{tabular}
\end{table}

\begin{table}[ht]
    \centering
    \caption{Grain property}
    \label{tab:grain_mat_prop}
    \begin{tabular}{ll}
    \hline
    Young's Modulus & $64 GPa$ \\
    Coefficient of Restitution & $0.66$\\
    Poisson's Ratio & $0.2$\\
    Coefficient of Friction & $0.84$ \\
    Density & $2600 kg/m^3$ \\
    Grain radius & 1.2 $\pm 0.24 mm$ \\
\hline
    \end{tabular}
\end{table}
\par
The material properties of the grains and impact disc are summarized in Table \ref{tab:grain_mat_prop}.
The Bennu sampling assessment \cite{Hoover_2024_lpsc} reports Young's modulus of 183.2 $\pm$ 44 GPa and hardness of 12.5 $\pm$ 3.56 GPa. 
For the simplicity of computation, the Young's Modulus is reduced to 64 GPa per previous work \cite{GROGER2003}. \par
The coefficient of friction is kept constant at 0.84 because the role of friction within the interaction of cohesion and gravity is beyond the scope of this study.
It is worth noting that friction and granular packing both affect the bulk properties of a granular material \cite{surfacefrictionPRE2018,fdistandfriction2001}.\par

\begin{figure}[ht]
\centering
\includegraphics[width=0.4\textwidth]{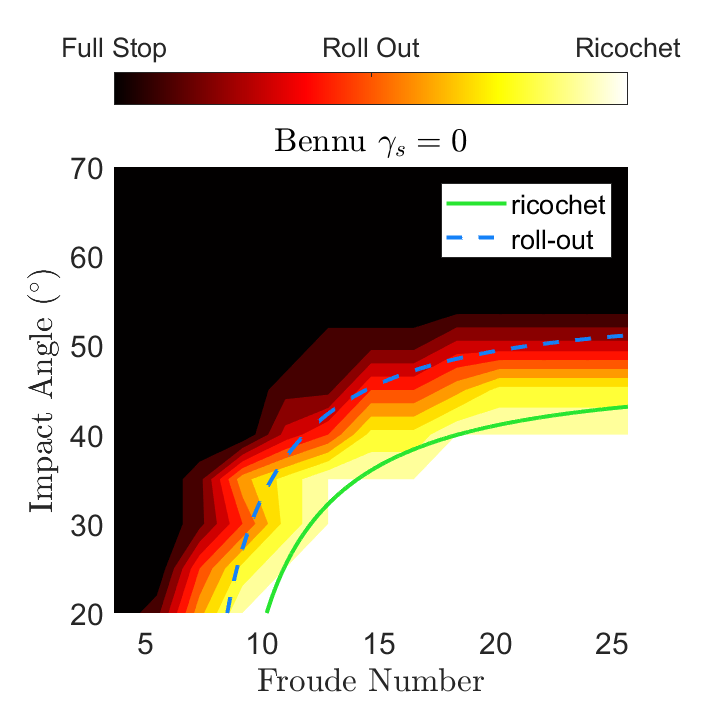}
\caption{Behavioral map of impacts on cohesion-less bed under Bennu gravity condition, using damping coefficient $\eta_{n0}$=500. At each Froude number and impact angle, the impact behavior is either full-stop (black) when the impactor remains in the initial crater without escaping it, roll-out (red) when it escapes the initial crater without loosing its contact with the surface particles, or ricochet (white) when it escapes the crater without further contact with the surface. The map shows an average of the behavior of the impactor from three different initial locations. The behavior roughly follows the fitted lines suggested by Wright et al. \cite{esteban2020}.}
\label{fig:band}
\end{figure}

Adopting our previous definition of impact outcomes categorization\cite{esteban2020, peter2023}, the behavior of the impactor is divided into either full-stop, ricochet, or roll-out; full-stop is to remain in the initial crater, ricochet is to escape it without further contact with the surface, and roll-out is to escape while maintaining constant contact with the surface.
To summarize the behavior, we use a behavior map that records the impactor behavior as black for full-stop, red for roll-out, and white for ricochet at the corresponding Froude number and impact angle.
For example, Figure \ref{fig:band} summarizes the simulation result of a cohesion-less bed under Bennu gravity using a damping coefficient $\eta_{n0}$=500.
The behavior matches the experimental result of cohesionless granular media, shown as solid and dashed lines \cite{esteban2020}.
\par

The cohesive interaction among neighboring particles alters the bulk material properties of the granular media and how it responds to an impact.
However, the link between cohesion of individual particles and the bulk material property of the particles is unclear due to the stochastic nature of contact in granular media.
In other words, granular media may have a different bulk cohesion even when the individual grains have the same cohesion based on the size ratio and the placement of the particles. One can think of a crystallized configuration of grains where all cohesive bonds are established and contribute to the overall cohesive strength as opposed to a natural packing where each grain only exhibits a few contact points.
In this study, cohesion is quantified as a bulk property of granular media that is resistant strength against detachment as a source of the resistance against bed deformation.\par
To transfer cohesion among individual grains into bulk cohesive property, we conducted tensile test simulations.
This treats the granular bed as a continuous ductile material in which a layer of particles is pulled upward while the bottom layer is fixed.
The selected layer of the particles that are moved up is located from the top to 38 mm deep, which is as deep as about 20 particles stacked up vertically.
With a set upward speed of 0.3 mm s$^{-1}$, the layer is displaced upward against gravity, friction ,and cohesive forces. This speed constitutes a nominal strain rate of 0.25 s$^{-1}$, which is a fairly small strain rate compared to the grain size and is chosen to avoid inertial effects.
\par
\begin{figure}[ht]
\centering
\includegraphics[width=0.4\textwidth]{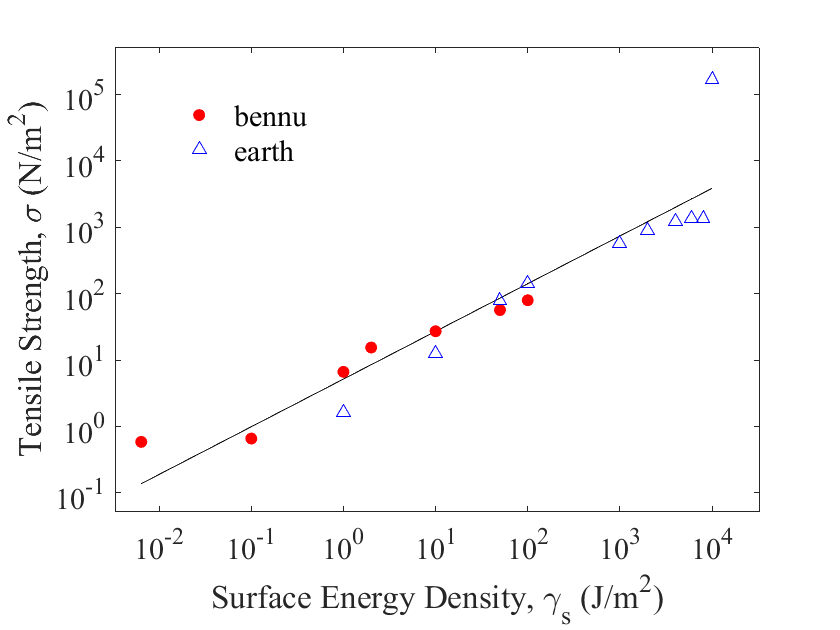}
\caption{Peak tensile strength $\sigma$ (N m$^{-2}$) at different cohesive surface energy density $\gamma_s$ (J m$^{-2}$) under Bennu (red circle) and Earth (blue triangle) gravity conditions. The granular bed is assigned different surface energy densities ranging from $\gamma_s$=0.01 J m$^{-2}$ to 10,000 J m$^{-2}$. The bottom layer of each bed is fixed, and tensile displacement is applied to the top layer. The resistive force experienced by the top layer is then recorded at each displacement. Among those forces, the peak resistive force is divided by the length of the bed and the nominal diameter of the disc to yield the tensile strength.}
\label{fig:cohesion_vs_gamma}
\end{figure}

The force experienced by the layer is then recorded at each discrete displacement, analogous to the stress-strain curve of a typical tensile test.
From the recorded force, the weight of the layer is subtracted to yield only the resistive force due to cohesion and surface and structural friction.
The vertical displacement of the layer, however, engages a negligible amount of shear that minimizes the effect of the surface and structural friction on resistive force.
Thus, the resistive force is approximated to a cohesive stress value by dividing it by the product of the span of the bed length and the nominal diameter of the grains.
The peak resistive stress against detachment continuously increases as the coefficient of cohesion increases from Bennu to Earth gravity case, as shown in Figure \ref{fig:cohesion_vs_gamma} with fitted line shown in Eq. \ref{eq:sig_vs_gam}.
\begin{equation}
\log_{10}(\sigma)=1.4354 \log_{10}(\gamma_s) + 1.4303
\label{eq:sig_vs_gam}
\end{equation}

As Sánchez et al. \cite{sanchez_scheeres_2014} pointed out, tensile strength is a function of grain size and thus Eq. \ref{eq:sig_vs_gam} is only applicable to this particular system.
The coefficient of cohesion values used in this study $\gamma_s$=0, 0.1, 0.2, 1.0 correspond to $\sigma$= 0, 0.0133, 0.1319, 26.9339 (Pa), respectively, which are in accordance with Sánchez et al. \cite{sanchez_scheeres_2014} but smaller than the nominal stress previously suggested \cite{Bennu_coh_sample_2024}.
\par

The timestep formula proposed by Burns et al. \cite{burnststep2019} requires
\begin{align}
\Delta t = \frac{\pi r}{\beta} \sqrt{ \frac{\rho}{G}}
\label{eq:timestep}
\end{align}
where \(r\) is the radius of the particle, \(\beta\) is an experimental value represented as \(\beta = 0.8766 + 0.163 \nu\) and \(G\) is the shear modulus of the grain which is related to Young's Modulus \(E\) and Poisson's ratio \(\nu\) as \(E=2G(1+\nu)\).
According to Eq. \ref{eq:timestep}, a critical time step of 3.668 ms is a stable timestep for this study, but it is kept as 1 $\mu$s for Earth and Moon gravity conditions and 5 $\mu$s for Bennu conditions for a conservative result at the cost of an affordable computation time. It is also worthwhile to mention that the ambient air pressure at high-speed impact affects the post-impact granular behavior \cite{ambientair2005}, but the air pressure is not considered in this study.\par

\section{Result and Discussion}
\subsection{Behavioral Map of DEM}
\begin{figure*}[ht]
\centering
{\includegraphics[width=\textwidth]{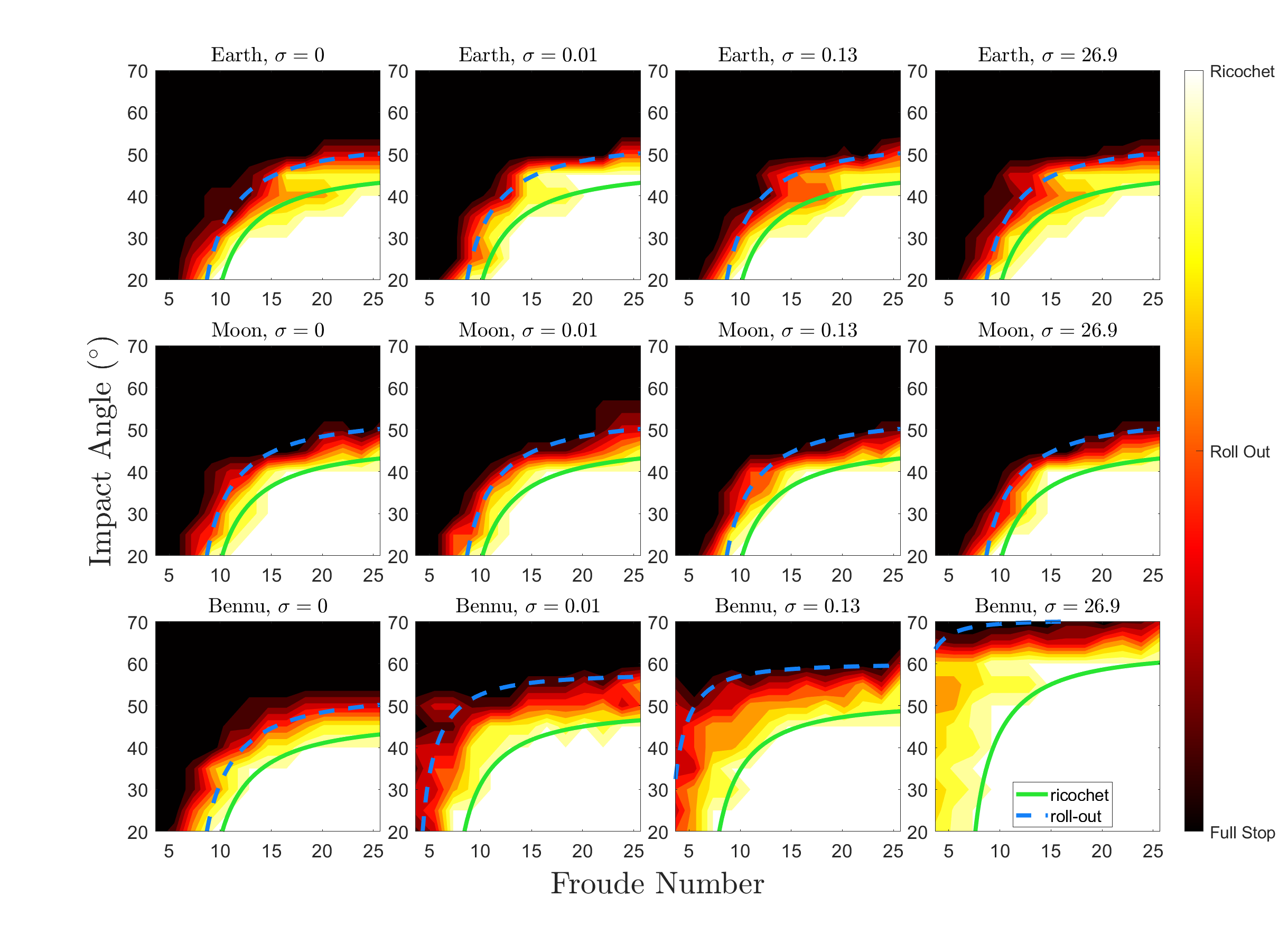}}
\caption{Behavioral map of impacts on cohesion-less granular bed (left column), cohesive bed with $\sigma$=0.01 Pa (second column from left), $\sigma$=0.1 Pa (third column from left), and $\sigma$=26.9 Pa (last column from left) at Bennu (bottom row), Moon (center row), and Earth (top row) gravity conditions. Impact behavior is represented as white for ricochet, red for roll-out, and black for full-stop. The difference in the behavior map for Bennu gravity conditions for different $\sigma$ values is larger than those for Moon or Earth conditions. The average packing fraction is $\phi$=0.812 (Earth) and 0.811 (Moon and Bennu). The fitted lines suggest the behavior difference of each map; between full-stop and roll-out is a solid blue line, and roll-out and ricochet a dashed green lines based on Eq. \ref{eq:generalform}.}
\label{fig:behavemap}
\end{figure*}

The behavior map of the impacts at three gravity conditions and four cohesion levels is shown in Figure \ref{fig:behavemap}.
The bulk cohesive strength, $\sigma$, increases from left to right columns marking 0, 0.01, 0.13, and 26.9 Pa.
The gravity magnitude increases from the bottom row to the top row, each row representing Bennu, Moon, and Earth gravity conditions, respectively, from the bottom.\par
The maps on the left column show the same overall behavior as the cohesionless cases \cite{peter2023}. Our use of the JKR model \cite{JKR1971} results in a slightly larger area of ricochet in Bennu than in Moon or Earth gravity conditions.
\par
The general shape of the curves of ricochet stays constant over different $\sigma$ values in Earth and Moon gravity conditions, but it expands horizontally and vertically as cohesion increases in Bennu gravity conditions.
For instance, the bottom row of behavior maps shows a decreasing number of full-stop outcomes at low Froude number and low impact angle as cohesion increases, which ultimately pushes this behavior out of our bounds of study at $\sigma$=26.9 Pa.

These tendencies across different gravity and cohesive conditions are explained with separation lines, similar to those suggested by Wright et al. \cite{esteban2020}.
We introduce a modified Froude number $\pi_5$ to incorporate cohesion as a source of inter-granular pressure along with gravity
\begin{equation}
\pi_5 =\frac{\rho v^2a}{\rho g a^2 +c_g \gamma_s}
\end{equation}
where $c_g$=100,000 for this system.
The dimensionless constant $c_g$ represents the effect of cohesion with respect to inter-particular friction, and thus depends on properties such as impactor radius, grain radius, and particle density, which is consistent with previously reported data \cite{tagsamsimul_2021}.
When cohesion is negligible, $\pi_5$ reduces to the Froude number.
The separation lines are in the form of

\begin{equation}
    \theta^2=A-\frac{B}{(\pi_5)^2}
    \label{eq:generalform}
\end{equation}
where between ricochet and roll out
\begin{equation}
    A_{rc}=0.0001540 \left(\frac{\gamma_s}{g}\right)^{0.8751}+0.65
        \label{eq:Arc}
\end{equation}

\begin{equation}
B_{rc}=55-\frac{26}{1+2440.6\left(\frac{g}{\gamma_s}\right)^{1.1292}}
\label{eq:Brc}
\end{equation}
and between roll out and full stop
\begin{equation}
A_{ro}=0.0017\left(\frac{\gamma_s}{g}\right)^{0.6128}+0.85
\label{eq:Aro}
\end{equation}
\begin{equation}
B_{r0}=55-\frac{54.9925}{1+828.8175\left(\frac{g}{\gamma_s}\right)^{1.0423}}
\label{eq:Bro}
\end{equation}

The area below the ricochet separation curve and the area between the two separation lines increase as cohesion increases in Bennu gravity condition.
Such expansions happen because a more cohesive bed forms a shallower crater and a smaller curtain.
Cohesion among the grains induces a stronger bed surface, thus making a smaller crater than on a cohesion-less bed surface with the same amount of impact energy.
However, the surface strength difference is only evident in micro-gravity conditions where cohesion is significant relative to the weight and friction-induced pressure.
\par
The critical angle above which only full-stop behavior appears is about 50$^\circ$ for Earth and Moon gravity conditions for all cohesion values.
The critical angle of Bennu gravity condition changes from 50$^\circ$ to 60$^\circ$ for $\sigma \le$0.2, and it is beyond the scope of the acquired data for $\sigma$=26.9 Pa.\par
The Earth maps at $\sigma$=0, 0.13, and 26.9 Pa show an unexpected row of roll-out cases at angle 40$^\circ$ from Froude number 15 to 20.
This pattern is due to the subsurface irregularity of one specific impact point that induced an unexpected full-stop behavior.
The pattern does not appear in the other gravity conditions due tothe  reorganization of pressure gradients among the grains under reduced weight.\par

\subsection{Reduced-order theoretical modeling}
Parallel to DEM, in which forces are calculated based on the individual grain interactions, the granular media can be considered as a continuous resistive medium that opposes the intrusive motion of foreign objects interacting with it. 
The approach is referred to as the Resistive Force Theory (RFT)  \cite{li2013terradynamics2013}. 
RFT is a framework that predicts the forces on objects moving through granular materials such as sand or soil. 
It assumes that the resistive force at each small surface element of the object depends only on the local properties of the granular medium and the object's velocity at that point.
The total resistive force is determined as the summation of these localized forces over the object's area.
RFT simplifies complex granular interactions by treating the medium as quasi-static while neglecting time-dependent effects.\par
This theory is widely used for designing locomotion in robots and understanding animal movement in granular environments  \cite{Maladen2011,ZhangT2014}. Later,  its physical foundation was explored extensively \cite{askari2016intrusion}.
Once the intruder velocity exceeds the quasi-static realm, Dynamic Resistive Force Theory \cite{agarwal2021surprising} defines the magnitude of resistive force.
We suggest an expansion of DRFT to include cohesion forces by adding a cohesive force term.
The total force on an intruding object into a cohesive and frictional granular medium can be written as

\begin{equation}
\begin{split}
    \vec{F}  &= (\vec{\tau}_{QS} + \vec{\tau}_{RD} + \vec{\tau}_c) A \\
    \vec{\tau}_{QS}&= \vec{\alpha} \left(\beta,\gamma \right) H(-\bar{z}) |\bar{z}| \\
    \vec{\tau}_{RD}&= \lambda \rho |V|^2 \hat{\mathbf{n}}\\
    \vec{\tau}_c&= c H(\hat{\mathbf{n}} \cdot \hat{\mathbf{v}}) \hat{\mathbf{n}}
\end{split}
\end{equation}
where \(\vec{\tau}_{QS}\) is quasi-static resistive traction,
\(\vec{\tau}_{RD}\) is inertial resistive traction,
\(\vec{\tau}_{c}\) is cohesive traction,
\(A\) is the surface integration of the intrusive object,
\(\vec{\alpha}(\beta,\gamma)\) is the resistive force coefficient,
\(\bar{z}\) is the depth of the center of mass of the intrusive object,
\(\textbf{H}\) is the Heaviside function,
\(z\) is the depth of the intrusive object,
\(\lambda\) is the shape factor of dynamic resistive force,
\(\rho\) is the critical particle density,
\(|V|\) is the normal component of the velocity to the element surface,
\(\hat{{\bf n}}\) is unit normal vector of the element and
\(\hat{{\bf v}}\) is unit velocity vector of the element. \par

The resistive force coefficient $\alpha$ depends on the granular bed properties such as volume fraction, friction, and density of the particles, the angle between resistive force and object orientation $\beta$, and the subsurface object motion direction $\gamma$.
The scalar fitting constant $\lambda$ is determined based on traction forces experienced by the object during an intrusive motion, which depends on the velocity.
$\alpha$ and $\lambda$ are acquired experimentally due to their dependence on environment-specific properties such as localized volume fraction or intrusive motion dynamics \cite{agarwal2021surprising,li2013terradynamics2013}.
\par

It has previously been shown that the Froude number scaling unifies the trajectories of oblique impacts into cohesionless granular media in different gravities \cite{peter2023}. Here, we examine a similar collapse of behavior in the presence of cohesion. The equations of motion under Earth's gravity can be written as
\begin{align}
[\alpha_z^o g z(t) + \lambda \rho [D_t z(t)]^2] A - mg = m D_t^2 z(t)
\label{origeom}
\end{align}
and at a reduced gravitational  condition
\begin{align}
[\alpha_z^o g^* z^*(t^*) + \lambda \rho [D_t z^{*}(t^*)]^2] A - mg^* = m D_t^2 z^*(t^*)
\label{stareom}
\end{align}
where $g$ is Earth gravitational acceleration, $g^*$ is the reduced gravitational acceleration of either Moon or Bennu, and $z(t)$ and $z^*(t^*)$ are the solutions of Eq. \ref{origeom} and \ref{stareom}.
Assuming that the cohesive traction is independently added as a  cohesive traction $c$, Eq. \ref{origeom} becomes
\begin{align}
[\alpha_z^o g z(t) + \lambda \rho [D_t z(t)]^2 + c] A - mg = m D_t^2 z(t).
\label{cohorigeom}
\end{align}
Substituting the variables with those of reduced gravity conditions, we obtain
\begin{align}
[\alpha_z^o g^* z^*(t^*) + \lambda \rho [D_{t^*} z^*(t^*)]^2 + c^*] A - mg^* = m D_{t^*}^2 z^*(t^*).
\label{cohstareom}
\end{align}
The linear time scaling is considered such that $t^*=t \sqrt{g/g^*}$. Then $D_{t*} = \sqrt{g^*/g} D_t$ and $D^2_{t*} = g^*/g D^2_t$.

\begin{figure}[ht]
\centering
\includegraphics[width=0.4\textwidth]{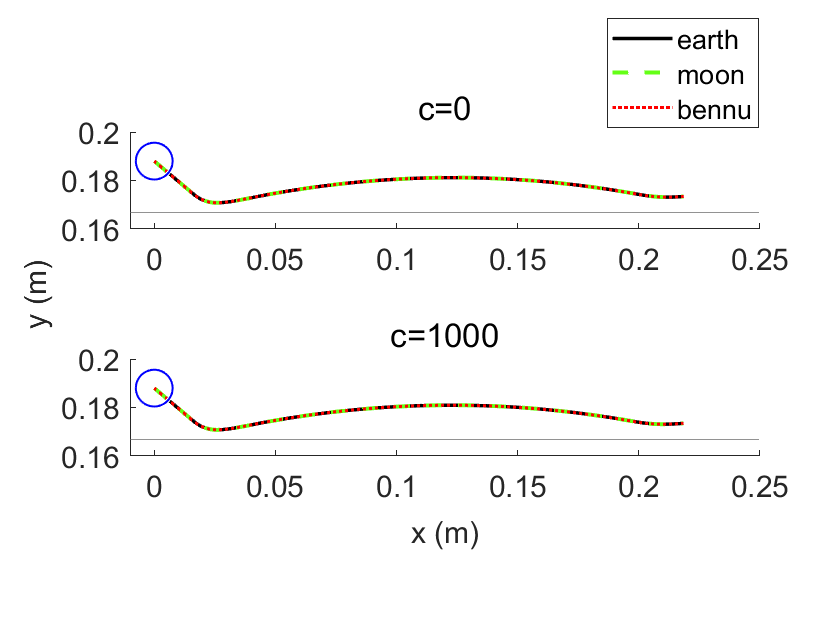}
\caption{Examples of matching impact trajectory based on Eq. \ref{cohstareom} with cohesion coefficient $c$=0 (top) and $c$=1000 (bottom). For both cases of $c$, the trajectories match among various gravity conditions of Earth (black solid line), Moon (green dashed line), and Bennu (red dotted line) when $c^*$ is scaled with $g/g^*$. The impact angle is 40 $^\circ$, and Froude number is 25.7.}
\label{fig:matlab_implementation}
\end{figure}

Then we multiply ${g}/{g^*}$ to both sides to get
\begin{align}
[\alpha_z^o g z^*(t^*) + \lambda \rho [D_t z^*(t^*)]^2 + c^*\frac{g}{g^*}] A - mg = m D_t^2 z^*(t^*)
\label{cohstareomreturned}
\end{align}
Thus, Eqs. \ref{cohorigeom} and \ref{cohstareomreturned} have the same format, indicating $z(t)$ and $z^*(t^*)$ have the same trajectory when cohesion coefficient $c^*$ is scaled with ${g}/{g^*}$.
Therefore, the trajectory of the impactor matches at different gravity conditions when the ratio of cohesion coefficient and gravity is kept constant.
The ratio is expressed by Bond number ${\rho g a^2}/{\gamma_s}$, where $a$ is the impact disc radius.
Figure \ref{fig:matlab_implementation} shows the visualization of a discrete numerical implementation of Eq. \ref{cohstareom} using MATLAB to present as a contour map.

\begin{figure}[ht]
\centering
{\includegraphics[width=0.4\textwidth]{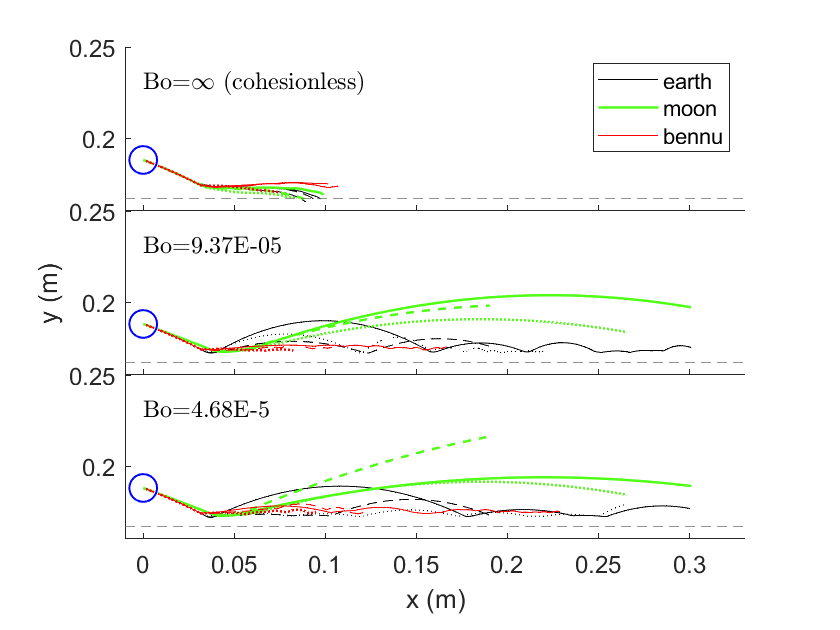}}
\caption{Impact trajectory of DEM at impact angle 20$^\circ$ and Froude number 5.50 when the cohesion of the bed is scaled with Bond number of $\infty$ (top), 9.37E-05 (center), and 4.68E-05 (bottom) under Earth (black), Moon (green) and Bennu (red) gravity conditions. Each of the three different styles of lines represents initial disc location. Unlike the MATLAB implementation of Eq. \ref{cohstareomreturned}, DEM trajectories diverge under different gravity conditions when the coefficient of cohesion is scaled with Bond number. The average local packing fraction is 0.93 for Earth and Moon and 0.90 for Bennu, calculated within 20 $\bar{r}$ from the impact point.}
\label{fig:DEM_traj}
\end{figure}

The DEM results, in contrast with the theoretical analysis of Eq. \ref{cohstareomreturned}, does not show a matching trajectory with the coefficient of cohesion scaled with Bond number, as shown in Figure \ref{fig:DEM_traj}.
The impactor shows a full-stop behavior onto a cohesionless bed under all gravity conditions.
However, the impactor ricochets under Moon and Earth gravity conditions while it shows a roll-out behavior under Bennu conditions when the coefficient of cohesion is scaled with the Bond number of 9.37E-05.
Contrary to the trend observed in Figure \ref{fig:behavemap}, the trajectories for Earth and Moon show more ricochet cases than Bennu at a higher Bond number condition. 
This is because the extreme cohesion among the particles causes them to have strong adhesion between those in contact, which ultimately results in a localized internal failure point as a surface void in the bed. 
Therefore, the scaling suggested by the theory through the additive contribution of cohesive forces and frictional and inertial forces is not verified by DEM numerical simulations. \par
One possible reason for the discrepancy between DEM results and the theoretical prediction is that neither of the experimental value $\alpha \left(\beta,\gamma \right)$ or the shape factor $\lambda$ is calibrated for this example.
Kerimoglu et al. \cite{kerimoglu_2024} argue that cohesive grains show a higher resistance than non-cohesive granular media, resulting in a larger $\alpha \left(\beta,\gamma \right)$.
But prediction of $\lambda$ under different cohesive condition is yet to be discussed.\par
We can consider a vertical plate and calculate $\alpha_{x,z}\left(\pi/2,0 \right)$ and $\lambda$ by DEM using a constant speed for the plate motion within the same bed used for our impact analyses and considering a coefficient of cohesion $\gamma_s$=0.2 for Bennu gravity condition and $\gamma_s$1.0 for Earth.
A block of particles is replaced with a plate at 0.05 m in the horizontal direction and from 0.12 m to 0.145 m in the vertical direction, setting the area of the plate as $A$=2.5 cm$^{2}$.
As the plate moves with a constant velocity within the bed, the contact force between particles and the plate is recorded as the total resistive force.\par
Then from the recorded force, $\alpha_{x,z}$ and $\lambda$ are obtained using Eq. 4 Agarwal et al. \cite{agarwal2021surprising}.
The quasi-static velocity is 0.51 mm s$^{-1}$ for Bennu gravity condition and 1.2 mm s$^{-1}$ for Earth to acquire $\alpha_{x,z}$.
To calculate $\lambda$, higher velocities are needed to enter the inertial regimes. These velocities are considered as 2.53, 25.3 and 253 mm s$^{-1}$ for Bennu gravity condition and 5, 1, 0.8, 0.6, 0.4 and 0.2 m s$^{-1}$ for Earth.
The resultant $\alpha_{x,z}$ and $\lambda$ are reported in Table \ref{tab:alpha_lambda}.\par
Compared to the cohesionless beds, $\alpha_x$ of the cohesive bed increases by 441\% for the Bennu gravity condition, whereas it decreases by 2.07\% for Earth.
Similarly, $\alpha_z$ increases by 611\% for cohesive Bennu gravity conditions but decreases by 4.5\% for cohesive Earth compared to cohesionless Bennu and Earth beds, respectively.
Each cohesive bed shows a larger $\lambda$ than its cohesionless counterpart for both gravity conditions -- by 2.16\% for Bennu and 4.50\% for Earth.\par
The slight decrease of $\alpha_{x,z}$ for cohesive bed compared to cohesion-less bed under Earth gravity conditions implies that the effect of cohesion $\gamma_s$ on both $\alpha_{x,z}$ and $\lambda$ depends on the gravity condition.
Our result seemingly disagrees with the previous work of Kerimoglu et al.  \cite{kerimoglu_2024}.
However, such difference is negligible compared to the increase of those values under Bennu gravity condition, implying coefficient of cohesion $\gamma_s$=1.0 is an insignificant cohesion value for Earth gravity to affect $\alpha_{x,z}$, whereas $\gamma_s$=0.2 greatly affects $\alpha_{x,z}$ for Bennu.
\par

\begin{table}[ht]
    \centering
    \caption{}
    \label{tab:alpha_lambda}
    \begin{tabular}{lllll}
    $g*$ & $\gamma_s$ & $\alpha_x$ [N cm$^{-3}$] & $\alpha_z$ [N cm$^{-3}$] & $\lambda$ \\
    \hline
    Earth & 0 & 1.12 & -4.75e-2 & 0.92 \\ 
    Earth & 1.0 & 1.10 & -3.09e-2 & 0.96 \\ 
    Bennu & 0 & 1.70e-5 & -3.23e-4 & 0.71 \\ 
    Bennu & 0.2 & 9.20e-5 & -2.30e-2 & 0.73 \\ 
\hline
    \end{tabular}
\end{table}
\par

The structure of the granular media and its grain arrangements give rise to the resistance forces experienced by the objects interacting with it. Even frictionless grains exhibit an overall coefficient of friction due to this configurational aspect   \cite{duran2001}. Cohesion acts in a similar manner. 
Cohesional forces serve as a source of intergranular pressure, which leads to an increased packing fraction as well as an increase in frictional forces between the grains.  An increase in pressure, will give rise the frictional forces by $\mu P$. This increase is not considered in the theoretical formula of the modified resistive force theory. This suggests that both coefficients of $\alpha$ and $\lambda$ must be modified in a consecutive theory accordingly in order to accommodate the increased pressure due to cohesion.\par

\section{Conclusion}
We examined the combined effect of cohesion and gravity in low-speed impact dynamics using numerical and theoretical analytical modeling and dimensionless parameters such as Froude number and Bond number to relate different gravitational conditions and cohesive strengths.
Presenting our results in behavior maps, we show that the effect of cohesion on impact trajectory intensifies in a lower gravity condition.
By modifying the Froude number to include cohesion, we found a simple curve that successfully matched ricochet and roll out lines seen in our simulations.  
However, we found that our simulations were not consistent with simple modifications to Dynamic Resistive Force Theory.\par
The inconsistency between the simulation results and the outcome of the suggested scaling analysis using Froude and Bond numbers alone implies a confounding between cohesion and friction.
Thus, it seems reasonable to define a new dimensionless number that explains the field parameter with cohesion for the purpose of predicting the impact behavior.
In future studies, the relationship between $\alpha \left(\beta,\gamma \right)$, $\lambda$, Froude and Bond number should be investigated to suggest a more reliable model using a scaling factor with well-defined friction and cohesion relationship.


\end{document}